\documentclass[camera]{jpaper}

\usepackage[nocompress]{cite}
\usepackage{algorithmic}
\usepackage{array}

\makeatletter
\let\MYcaption\@makecaption
\makeatother

\usepackage[font=footnotesize]{subcaption}

\makeatletter
\let\@makecaption\MYcaption
\makeatother

\usepackage{fixltx2e}
\usepackage{dblfloatfix}
\usepackage[nolessnomore, italic]{mathastext}
\usepackage[T1]{fontenc}
\usepackage[usenames,dvipsnames,svgnames,table]{xcolor}
\usepackage{multirow}
\usepackage{hhline}
\usepackage[normalem]{ulem}
\usepackage{enumitem}
\usepackage{setspace}
\usepackage{indentfirst}
\usepackage{footmisc}
\usepackage{fancyhdr}
\usepackage[us,12hr]{datetime}
\usepackage[keeplastbox]{flushend}
\usepackage[hidelinks]{hyperref}
\usepackage{authblk}
\usepackage{wrapfig}

\newif\ifcameraready
\camerareadytrue

\newcommand{\versionnum}[0]{5.1}

\ifcameraready
\else
\fi

\ifcameraready
  \newcommand{\todo}[1][]{}
  \newcommand{\chI}[1]{#1}
  \newcommand{\chII}[1]{#1}
\else
  \newcommand{\todo}[1][]{\textbf{\fcolorbox{black}{red}{\color{white}{TODO}}} \underline{$\overline{\hbox{\emph{#1}}}$}}
  \newcommand{\chI}[1]{{\color{MidnightBlue} #1}}
  \newcommand{\chII}[1]{{\color{BrickRed} #1}}
\fi

\title{
    Heterogeneous-Reliability Memory:\\
    Exploiting Application-Level Memory Error Tolerance
}

\author{Yixin Luo$^1$\qquad%
Sriram Govindan$^2$\qquad%
Bikash Sharma$^{3,2}$\qquad%
Mark Santaniello$^{3,2}$\qquad%
Justin Meza$^{3,1}$%
\vspace{2pt}\\%
Aman Kansal$^2$\qquad%
Jie Liu$^2$\qquad%
Badriddine Khessib$^2$\qquad%
Kushagra Vaid$^2$\qquad%
Onur Mutlu$^{4,1}$}
\affil{\em $^1$Carnegie Mellon University\qquad
$^2$Microsoft Corporation\qquad
$^3$Facebook\qquad
$^{4}$ETH Z\"urich%
}

\begin{document}
\maketitle

\pagenumbering{arabic}



\begin{abstract}

This paper summarizes our work on characterizing application memory error
vulnerability to optimize datacenter cost via Heterogeneous-Reliability
Memory (HRM), which was published in DSN 2014~\cite{luodsn14}, and examines
the work's significance and future potential.
Memory devices represent a key component of datacenter total cost of ownership
(TCO), and techniques used to reduce errors that occur on these devices increase
this cost.  Existing approaches to providing reliability for memory devices
pessimistically treat all data as equally vulnerable to memory errors.  Our key
insight is that there exists a diverse spectrum of tolerance to memory errors in
new data-intensive applications, and that traditional one-size-fits-all memory
reliability techniques are inefficient in terms of cost.  For example, we found
that while traditional error protection increases memory system cost by 12.5\%,
some applications can achieve 99.00\% availability on a single server with a
large number of memory errors without any error protection.  This presents
an opportunity to greatly reduce server hardware cost by provisioning the right
amount of memory reliability for different applications.

Toward this end, in our DSN 2014 paper~\cite{luodsn14}, we make three main contributions to enable
highly-reliable servers at low datacenter cost.  First, we develop a new
methodology to quantify the tolerance of applications to memory errors.
Second, using our methodology, we perform a case study of three new
data-intensive workloads (an interactive web search application, an in-memory
key--value store, and a graph mining framework) to identify new insights into
the nature of application memory error vulnerability.  Third, based on our
insights, we propose several new hardware/software
heterogeneous-reliability memory system designs to lower datacenter cost while
achieving high reliability and discuss their trade-offs.  We show that our new
techniques can reduce server hardware cost by 4.7\% while achieving 99.90\%
single server availability.

\chI{We believe the notion of HRM opens up a sea of opportunities in
optimizing memory system and overall system cost, reliability, efficiency, and
performance in a manner that is aware of applications' tolerance to memory
errors. \chII{Thus,} our paper just scratches the surface of a large HRM exploration
space, which we hope future works will undertake in various novel ways, in a
wide variety of systems, ranging from datacenters to mobile and embedded
systems.}

\end{abstract}



\section{Introduction}

\chI{A warehouse-scale datacenter consists} of many thousands of machines running a
diverse set of applications, \chII{and comprises} the foundation of the modern
web~\cite{verma.eurosys15, Barroso09}.  While \chI{such} datacenters are vital to the operation of
companies such as Facebook, Google, Microsoft, and Yahoo!, reducing the cost of
such large-scale deployments of machines poses a significant challenge to
these and other companies.  Recently, the need for reduced datacenter cost has
driven companies to examine more energy-efficient server
designs~\cite{Elnozahy03} and build their datacenter installations in cold
environments to reduce cooling costs~\cite{Jobin12, Grundberg11} or use built-in
power plants to reduce electricity supply costs~\cite{Riekstin13}.

There are two main components of the \emph{total cost of ownership} (TCO) of a
datacenter~\cite{Barroso09}:  (1) capital costs (those associated with server
hardware) and (2) operational costs (those associated with providing
electricity and cooling).  Recent studies have shown that capital costs can
account for the majority (e.g., around 57\% in~\cite{Barroso09}) of datacenter
TCO, and thus represent the main impediment for reducing datacenter TCO.  In
addition, this component of datacenter TCO is only expected to increase going
forward as companies adopt more efficient cooling and power supply techniques.

Of the dominant component of datacenter TCO (capital costs associated with
server hardware), the cost of server processors and memory represents the key
component---around 60\% in modern servers~\cite{Kozyrakis10}.  Furthermore, the
cost of the memory in today's servers is comparable to that of the processors\chI{~\cite{Kozyrakis10}},
and is likely to exceed processor cost for data-intensive applications such as
web search and social media services, which use in-memory caching to improve
response time~\cite{yue.twitter16, Nishtala2013, huang.sosp13, ousterhout.acm11,
ongaro.sosp11} (e.g., a popular key--value store, Memcached, has been used at
Google and Facebook~\cite{Nishtala2013, huang.sosp13} for this purpose).

Exacerbating the cost of memory in modern servers is the use of memory devices
(such as dynamic random access memory, or DRAM) that provide error detection and
correction.  This cost arises from two components: (1) quality assurance testing
performed by memory vendors to ensure devices sold to customers are of a high
enough caliber
and (2) additional memory capacity for error detection and correction.  Device
testing has been shown to account for an increasing fraction of the cost of
memory for DRAM~\cite{dram-test-cost, ddrtester}.  The cost of additional memory
capacity, on the other hand, depends on the technique used to provide error
detection and correction.

Table~\ref{tbl:mem-rel-cost} compares several common memory error detection and
correction techniques in terms of which types of errors they are able to
detect/correct and the additional amount of capacity/logic they require (which,
for DRAM devices, whose design is fiercely cost-driven\chI{~\cite{mutlu.imw13, mutlu.date17}}, is proportional to
cost).  Techniques range from the relatively low-cost (and widely employed)
parity, SEC-DED (single error correction, double error detection), Chipkill~\cite{Dell97}, 
and DEC-TED (double error correction, triple error detection), all of which use different \emph{error-correcting codes}
(ECC) to detect and correct a small number of bits or chip errors, to the more
expensive RAIM~\cite{Meaney12} and Mirroring~\cite{memory-mirroring} techniques that replicate some (or all) of memory
to tolerate the failure of an entire DRAM dual in-line memory module (DIMM). The
additional cost of memory with high error-tolerance can be significant (e.g.,
12.5\% of the total memory capacity for SEC-DED and Chipkill, and as high as 125\% for Mirroring).

\begin{table}[h]
  \centering
  \vspace{5pt}
  \footnotesize
  \tabcolsep=0.1em
  \caption{Memory error detection and correction techniques.  ``$X/Y\ Z$''
  means a technique can detect/correct $X$ out of every $Y$ failures of $Z$.
  \chI{$n$ \chII{represents the parity of any \emph{odd} number of bits
  between 1 and 63}.}
  Adapted from~\cite{luodsn14}.}
  \label{tbl:mem-rel-cost}
  \begin{tabular}{lrlccc}
    \toprule
              Technique & Error Detection & (Correction)  & Added Capacity &
              Added Logic \\
    \midrule
                 Parity & \chII{$n$}/64 bits  & (None)        & \enspace\enspace1.6\%         & Low \\
                SEC-DED & 2/64 bits       & (1/64 bits)   & \enspace12.5\%         & Low \\
                DEC-TED & 3/64 bits       & (2/64 bits)   & \enspace23.4\%         & Low \\
 Chipkill~\cite{Dell97} & 2/8 chips       & (1/8 chips)   & \enspace12.5\%         & High \\
   RAIM~\cite{Meaney12} & 1/5 modules     & (1/5 modules) & \enspace40.6\%         & High \\
Mirroring~\cite{memory-mirroring} & 2/8 chips & (1/2 modules) & 125.0\%      & Low \\
    \bottomrule
  \end{tabular}
\end{table}

Yet even with well-tested and error-tolerant memory devices, recent studies
from the field have observed a rising rate of memory error
occurrences\chI{~\cite{Schroeder09,Sridharan2012,hwang.asplos12,meza.dsn15,kim.isca14,mutlu.date17,mutlu.superfi14}}.  This trend presents an
increasing challenge for ensuring high performance and high reliability in
future systems, as memory errors can be detrimental to both.
In terms of performance, existing error detection and correction techniques
incur a slowdown on each memory access due to their additional
circuitry~\cite{hwang.asplos12,Li10} and up to an additional 10\% slowdown due to
techniques that operate DRAM at a slower speed to reduce the chances of random
bit flips due to electrical interference in higher-density devices that pack
more and more cells per square nanometer~\cite{jeff_micro10}.  In addition,
whenever an error is detected or corrected on modern hardware, the processor
raises an interrupt that must be serviced by the system firmware (e.g., BIOS),
incurring up to 100~$\mu$s latency---roughly 2000$\times$ \chI{the latency of} a typical
50~ns memory access latency~\cite{jedec}---leading to unpredictable
slowdowns \chI{and sometimes even system hangs~\cite{meza.dsn15}}.

In terms of reliability, memory errors can cause an application to slow down, \chI{hang,}
crash, or produce incorrect results~\cite{Fiala2012}.  Software-level
techniques such as the retirement of regions of memory with
errors\chI{~\cite{PFA, mcelog, Tang2006, hwang.asplos12,meza.dsn15}} have been proposed to
reduce the \chI{occurrence} of memory error correction \chI{events} and prevent correctable errors from
turning into uncorrectable errors over time.  Hardware-level techniques, such as
those listed in Table~\ref{tbl:mem-rel-cost}, are used to detect and correct
errors without software intervention (but \chI{\emph{with}} additional hardware cost).  All
of these techniques are applied \chI{\emph{homogeneously}} to memory systems in a
\chI{\emph{one-size-fits-all}} manner.



Our goal in our DSN 2014 paper~\cite{luodsn14} is to \chI{(1)}~understand how tolerant
different data-intensive applications \chI{and different memory regions of each application} are to memory errors,
and \chI{(2)}~design a new memory system organization that matches
hardware reliability to \chI{the error tolerance of the application and the memory region} in order to
reduce system cost.
The {\bf main idea} of our approach
is to classify applications \chI{and memory regions} based on their memory error tolerance, and map
applications \chI{and memory regions} to {\em heterogeneous-reliability} memory \chII{(HRM)} system designs managed
cooperatively between hardware and software to reduce system cost.  We make the
following {\bf \textit{contributions:}}

\begin{enumerate}

  \item A new {\bf methodology} to quantify the tolerance of applications \chI{and their memory regions} to
    memory errors.  Our approach measures the effect of memory errors on
    application correctness and quantifies an application's ability to mask or
    recover from memory errors.

  \item A comprehensive {\bf characterization} of the memory error tolerance of
    three data-intensive workloads:  an interactive web search
    application~\cite{luodsn14, Reddi10}, an in-memory key--value
    store~\cite{luodsn14, memcached}, and a graph mining
    framework~\cite{luodsn14, graphlab}.  We find that there exists an order of
    magnitude difference in memory error tolerance across these three
    applications.
    \chI{We also find that there exists an order of magnitude difference in
    memory error tolerance across different memory regions of each application.}

  \item An {\bf exploration} of the design space of \chI{a family of} new memory system
    organizations, called \emph{heterogeneous-reliability memory}, which
    combines
    a heterogeneous mix of reliability techniques that leverage application \chI{and memory region}
    error tolerance to reduce system cost.  We show that \chI{an example use of} our techniques
    \chII{reduces} server hardware cost by 4.7\%, while achieving 99.90\% single server
    availability, \chI{based on a preliminary evaluation of an example HRM system}.
    
\end{enumerate}


\section{Characterizing Memory Error Tolerance}
\label{sec:char}

We characterize three \chI{commonly-used} data-intensive applications to
quantify their tolerance to memory errors:
\begin{itemize}[leftmargin=1.2em]
\item{\bf \em
WebSearch}~\cite{Reddi10}, an interactive web search application,

\item{\bf \em
Memcached}~\cite{memcached}, an in-memory key-value store, and

\item{\bf \em
GraphLab}~\cite{graphlab}, a graph mining framework.
\end{itemize}
We run these three applications
in \emph{real production systems}, and sample hundreds to tens of thousands of unique
memory addresses for each application.

\subsection{Characterization Methodology}

To understand how tolerant
different data-intensive applications are to memory errors, our
characterization consists of three components:
(1) characterizing the outcomes of memory errors on an application based on how
they propagate through an application's code and data, (2) characterizing how
safe or unsafe it is for memory errors to occur in different regions of an
application's data, and (3) determining how amenable an application's data is
to recovery in the event of an error.  We describe the implementation of each component
in detail in Sections~III and IV of our DSN 2014 paper~\cite{luodsn14}.

\begin{figure}[h]
\centering
\includegraphics[trim=0mm 210mm 110mm 0mm,clip,width=0.68\linewidth]{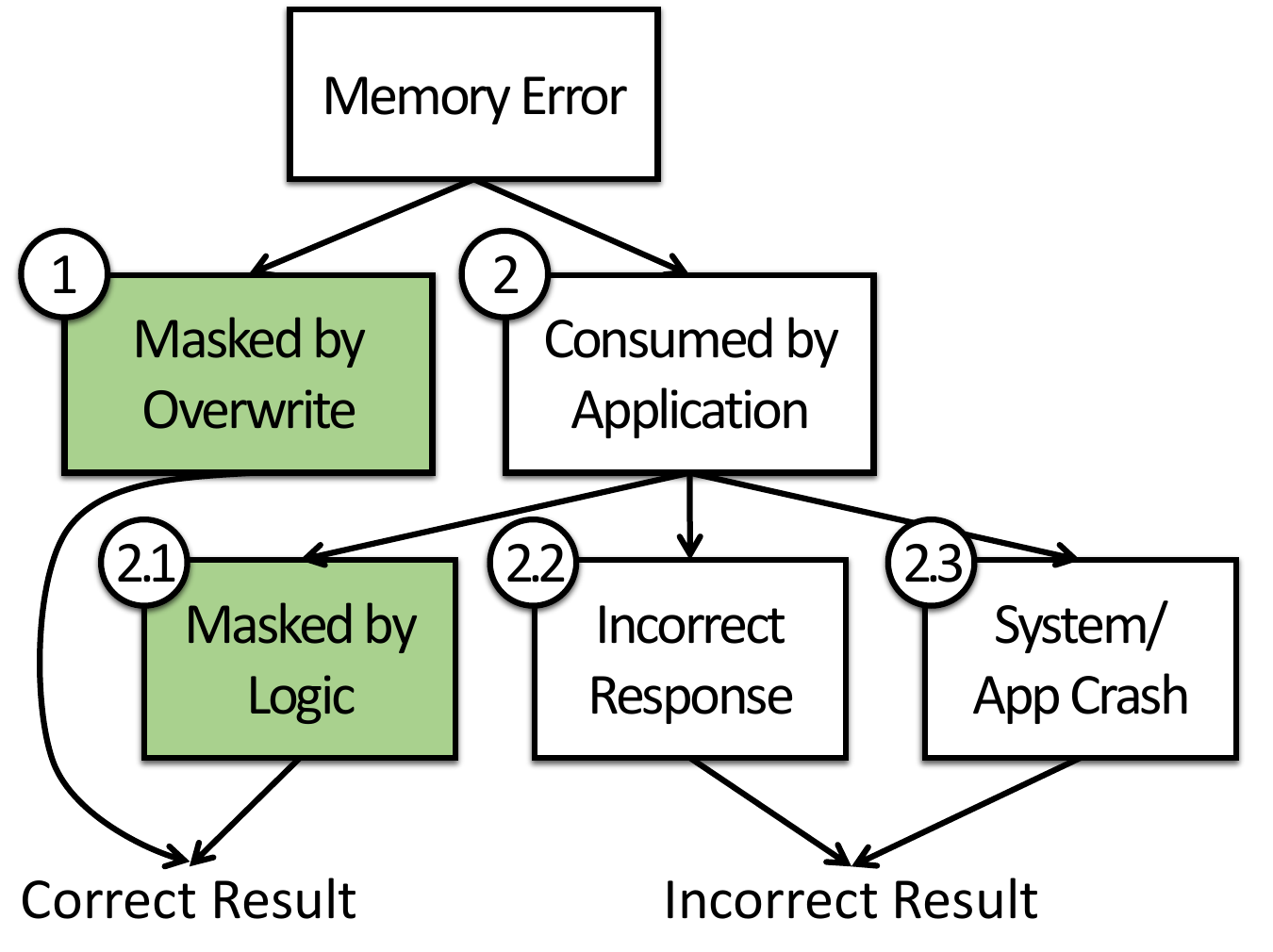}
\caption{\chI{Memory error outcomes. Reproduced from \cite{luodsn14}.}}
\label{fig:mem-err-outcome}
\end{figure}

\chI{We characterize an application's vulnerability to a memory error based on its
behavior after a memory error is introduced (we assume for the moment that no
error detection or correction is being performed).
Figure~\ref{fig:mem-err-outcome} shows a taxonomy of memory error outcomes.  Our
taxonomy is mutually exclusive (no two outcomes occur simultaneously) and
exhaustive (it captures all possible outcomes).  At a high level, a memory
error may be either (1) masked by an overwrite, in which case it is never
detected and causes no change in application behavior; or (2) consumed by the
application.  In the case that an error is consumed by the application, it may
either (2.1) be masked by application logic, in which case it is never detected
and causes no change in application behavior; (2.2) cause the application to
generate an incorrect response; or (2.3) cause the application or system to
crash.}

\chI{When we refer to the tolerance of an application to memory errors, we mean the
likelihood \chII{of} an error occurring in some data results in outcomes (1) or
(2.1).  Conversely, when we refer to the vulnerability of an application to
memory errors, we mean the likelihood \chII{of} an error occurring in some data
results in outcomes (2.2) or (2.3).}

We have three design goals when implementing our methodology for quantifying
application memory error tolerance.  First, due to the sporadic and inconsistent
nature of memory errors in the field\chI{~\cite{Schroeder09, meza.dsn15, Sridharan2013,
sridharan.asplos15, Sridharan2012,
kim.isca14, liu2013experimental, khan.sigmetrics14, qureshi2015avatar}}, we want to
design a framework that emulates the occurrence of a memory error in an
application's data in a {\bf \em controlled} manner.  Second, we want an {\bf
\em efficient} way to measure how an application accesses its data.  Third, we
want our framework to be easily {\bf \em adaptable} to other workloads or
system configurations. 

\begin{figure}[h]
\centering
\includegraphics[trim=8mm 110mm 210mm 5mm,clip,width=0.38\linewidth]{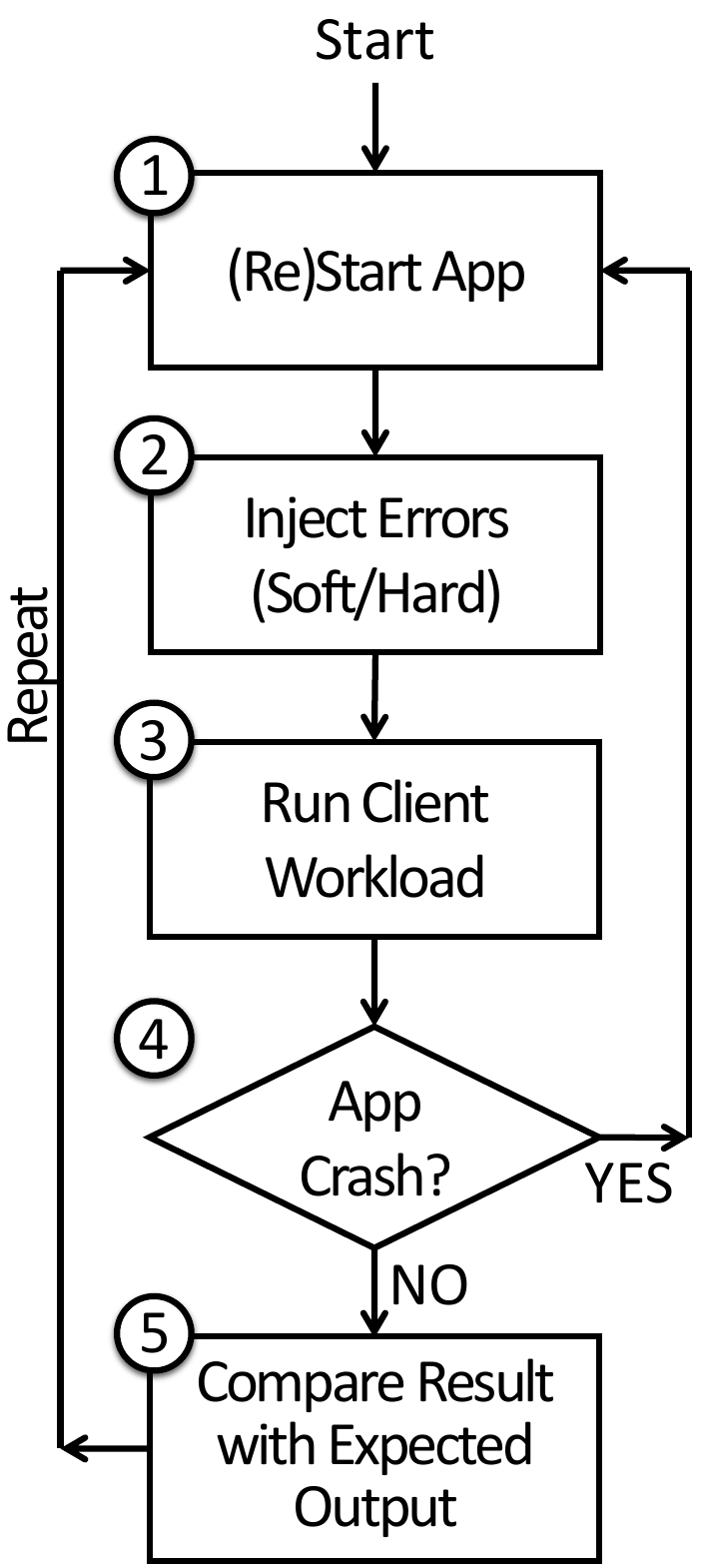}
\caption{Memory error emulation framework. Reproduced from \cite{luodsn14}.}
\label{fig:err-inj-framework}
\end{figure}

Figure~\ref{fig:err-inj-framework} shows a flow diagram illustrating the five
steps involved in our error emulation framework.  We assume that the application
under examination has already been run outside of the framework and its expected
output \chI{without any memory errors} has been recorded.  The framework proceeds as follows.  (1)~We start the
application under the error injection framework. Our memory
error emulation framework is described in Section~IV of our DSN 2014 paper~\cite{luodsn14}.  (2) We use
software debuggers\footnote{{\tt WinDBG}~\cite{Windbg} in Windows and {\tt
GDB}~\cite{Gdb02} in Linux.} to inject the desired number and types of memory errors.
(3)~We initiate the connection of a client and start executing the desired
workload.  (4)~Throughout the course of the application's execution, we check to
see if the machine has crashed; if it has, we log this outcome and proceed to
step (1) to begin testing once again.  (5)~If the application finishes its
workload, we check to see if its output matches the expected results; if the
output does not match the expected results, we log this outcome and proceed to
step (1) to test again.
\chI{Each run injects a particular pattern of errors into the application.
We can run this framework as many times as needed to test an application
with \chII{different patterns} of injected errors.}

There are two main types of memory errors: (1) {\em soft} or {\em transient}
errors and (2) {\em hard} or {\em recurring} errors.\footnote{\chI{Recent}
studies\chII{~\cite{khan.sigmetrics14, kim.isca14, liu2013experimental,
qureshi2015avatar,khan.dsn16, khan.micro17}}
examined the effects of
\emph{intermittent} and \emph{access-pattern dependent} errors, which are
increasingly common as DRAM technology scales down to smaller technology
nodes\chI{~\cite{mutlu.date17}}.} Soft memory errors occur at random due to charged particle emissions
from chip packaging or the atmosphere~\cite{May1979}. Hard memory errors may
occur from physical device defects or wearout~\cite{hwang.asplos12, Sridharan2012,
Schroeder09}, and are influenced by environmental factors such as humidity,
temperature, and utilization~\cite{Schroeder09, Sridharan2013, Siddiqua2013}.
Hard errors typically affect multiple bits (for example, large memory regions
and entire DRAM chips have been shown to fail~\cite{hwang.asplos12, Sridharan2012,
Sridharan2013}). Our characterization covers single-bit soft and hard errors.
\chI{For a detailed background on DRAM, we refer the reader to prior 
works~\cite{kim.hpca10, kim.micro10, SALP, TL-DRAM, lee.hpca15,
hassan.hpca16, liu2012raidr, chang.hpca14, chang.hpca16,RowClone, chang.sigmetrics17,
lee.sigmetrics17, seshadri.micro17, liu2013experimental, chang.sigmetrics16, hassan.hpca17, kim.cal15,
lee.pact15, lee.taco16, kim.isca14, patel.isca17, kim.hpca18}.}



\subsection{Key Findings}

We summarize two of the most important
findings from our characterization below.
We briefly list four other findings in Section~\ref{sec:char:other}, and
describe all six of our findings in detail in Section~V-B of our
DSN 2014 paper~\cite{luodsn14}.

\begin{figure}[h]
  \centering
\begin{subfigure}[b]{0.49\linewidth}
\includegraphics[trim=0mm 170mm 0mm 0mm,clip,width=\linewidth]{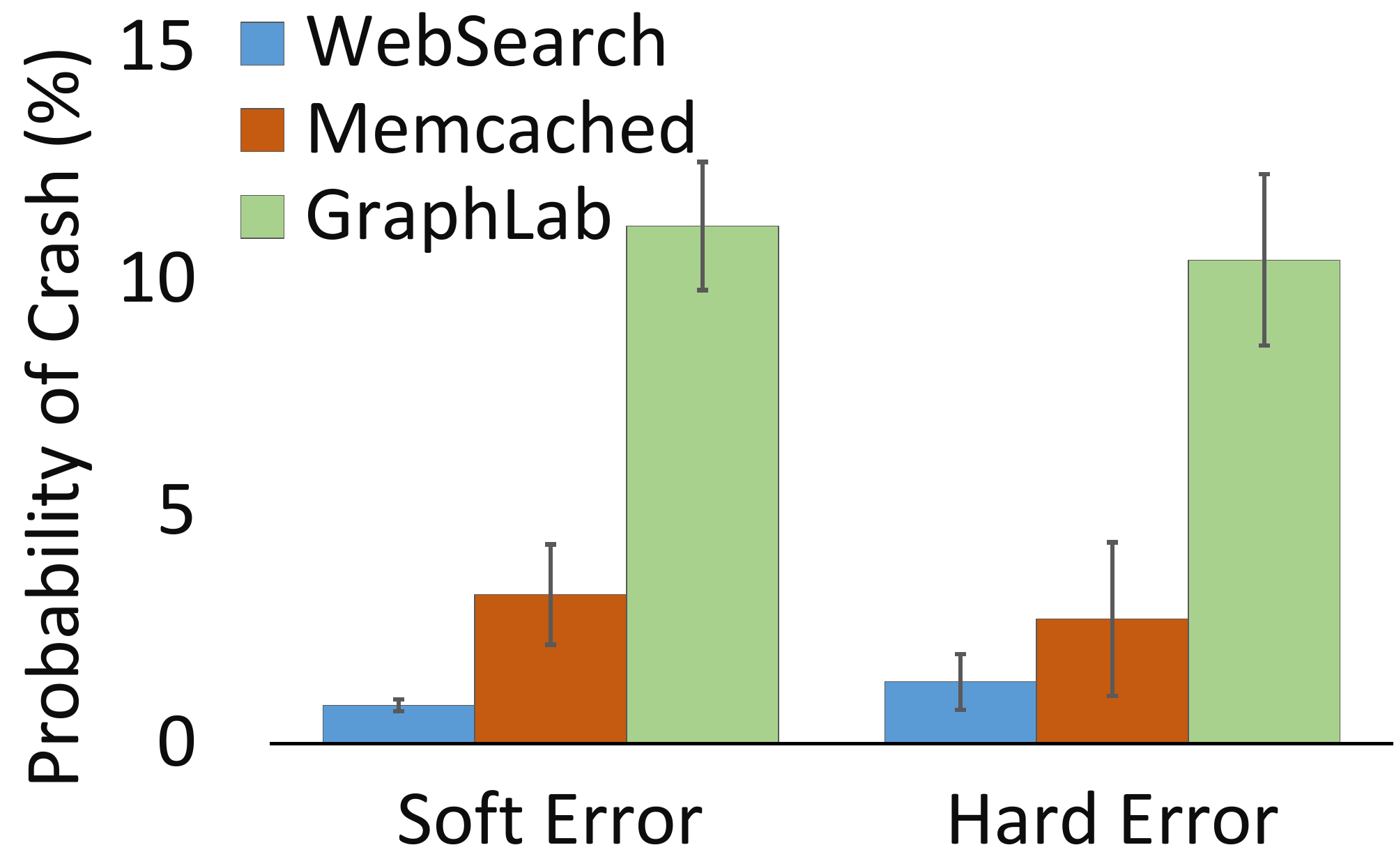}
\caption{Application vulnerability}
\end{subfigure}\hfill
\begin{subfigure}[b]{0.49\linewidth}
\includegraphics[trim=0mm 170mm 0mm 0mm,clip,width=\linewidth]{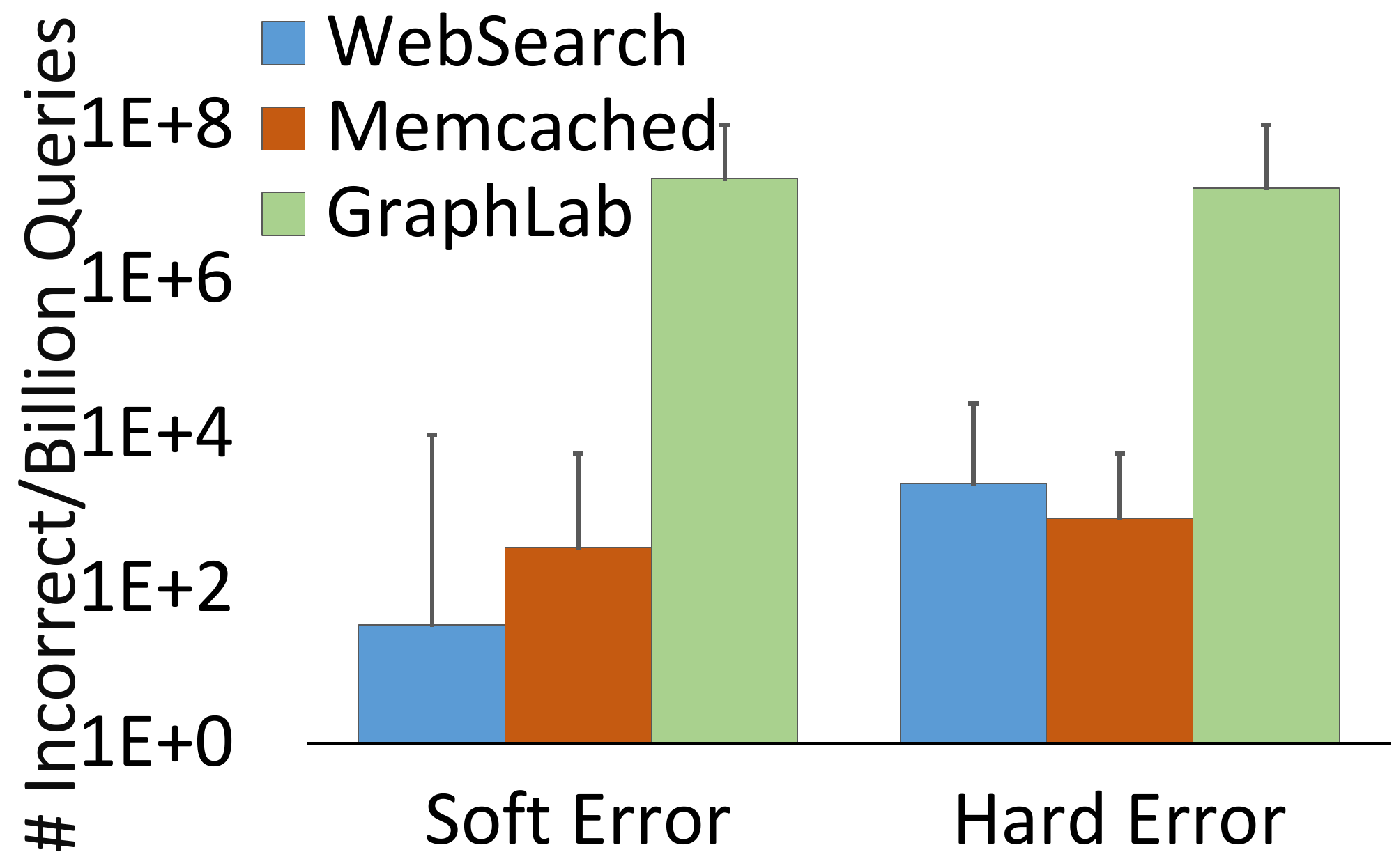}
\caption{Application incorrectness}
\end{subfigure}
  \caption{Inter-application variations in vulnerability to single-bit soft and
  hard memory errors for the three applications in terms of (a) probability of crash and (b) frequency of incorrect results. Reproduced from~\cite{luodsn14}.}
  \label{fig:inter-app-vulnerability}
\end{figure}

{\bf {Finding 1: Error Tolerance Varies Across Applications.}}
Figure~\ref{fig:inter-app-vulnerability}(a) plots the probability of each
of the \chI{evaluated} three applications crashing due to the occurrence of single-bit soft or
hard errors in their memory (we call this \emph{application-level memory error
vulnerability}).  \chI{For cases where the application \chII{does} not crash,}
Figure~\ref{fig:inter-app-vulnerability}(b) plots the rate of incorrect results
per billion application queries under the same conditions. We draw two key
observations from these results.

First, there exists a significant variance in vulnerability among the three
applications both in terms of crash probability and in terms of incorrect
result rate, which varies by up to six orders of magnitude.  Second, these
characteristics may differ depending on whether errors are soft or hard (for
example, the number of incorrect results for WebSearch differs by over two
orders of magnitude between soft and hard errors, \chI{with hard errors being more problematic}).  We therefore conclude that
{\em memory reliability techniques that treat all applications similarly are
inefficient because there exists significant variation in error tolerance among
applications.}

\begin{table*}[b!]
  \centering
  \vspace{5pt}
  \footnotesize
  \tabcolsep=0.1em
  \caption{Heterogeneous reliability design dimensions, \chI{example}
  techniques, and their potential benefits and trade-offs. \chI{Adapted} from~\cite{luodsn14}.}
  \label{tbl:design-space}
  \begin{tabular}{m{2.5cm}lll}
    \toprule
    Design dimension & Technique & Benefits & Trade-offs \\
    \midrule
    \multirow{6}{\linewidth}{\chI{Example hardware} techniques} 
    & No detection/correction & No associated overheads (low cost) & Unpredictable crashes and silent data corruption \\
    & Parity & Relatively low cost with detection capability & No hardware correction capability \\
    & SEC-DED/DEC-TED & Tolerate common single-/double-bit errors & Increased cost and memory access latency \\
    & Chipkill~\cite{Dell97} & Tolerate single-DRAM-chip errors & Increased cost and memory access latency \\
    & Mirroring~\cite{memory-mirroring} & Tolerate memory module failure & 100\% capacity overhead \\
    & Less-Tested DRAM & Saved testing cost during manufacturing & Increased error rates \\
    \midrule
    \multirow{5}{\linewidth}{\chI{Example software} responses}
    & Consume errors in application & Simple, no performance overhead & Unpredictable crashes and data corruption \\
    & Automatically restart application & Can prevent unpredictable application behavior & May make little progress if error is frequent \\
    & Retire memory pages & Low overhead, effective for repeating errors & Reduces memory space (usually very little) \\
    & Conditionally consume errors & Flexible, software vulnerability-aware & Memory management overhead to make decision \\
    & Software correction & Tolerates detectable memory errors & Usually has performance overheads \\
    \midrule
    \multirow{6}{\linewidth}{Usage granularity} 
    & Physical machine & Simple, uniform usage across memory space & Costly depending on technique used \\
    & Virtual machine & More fine-grained, flexible management & Host OS is still vulnerable to memory errors \\
    & Application & Manageable by the OS & Does not leverage different region tolerance \\
    & Memory region & Manageable by the OS & Does not leverage different page tolerance \\
    & Memory page & Manageable by the OS & Does not leverage different data object tolerance\\
    & Cache line & Most fine-grained management & Large management overhead; software changes \\
    \bottomrule
  \end{tabular}
\end{table*}

\begin{figure}[h]
  \centering
\begin{subfigure}[b]{0.495\linewidth}
\includegraphics[trim=0mm 150mm 0mm 0mm,clip,width=\linewidth]{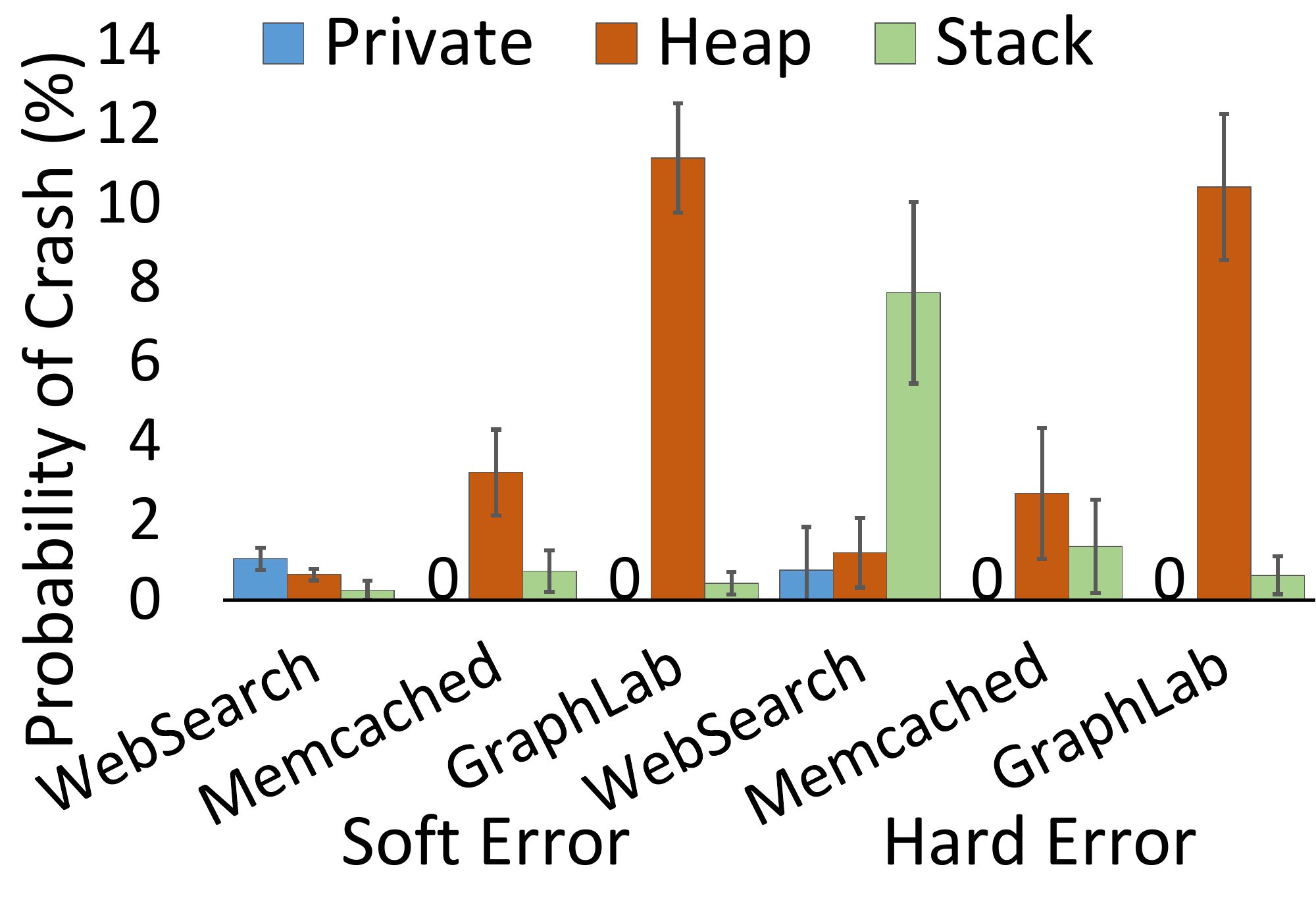}
\caption{Memory region vulnerability}
\end{subfigure}\hfill
\begin{subfigure}[b]{0.495\linewidth}
\includegraphics[trim=0mm 150mm 0mm 0mm,clip,width=\linewidth]{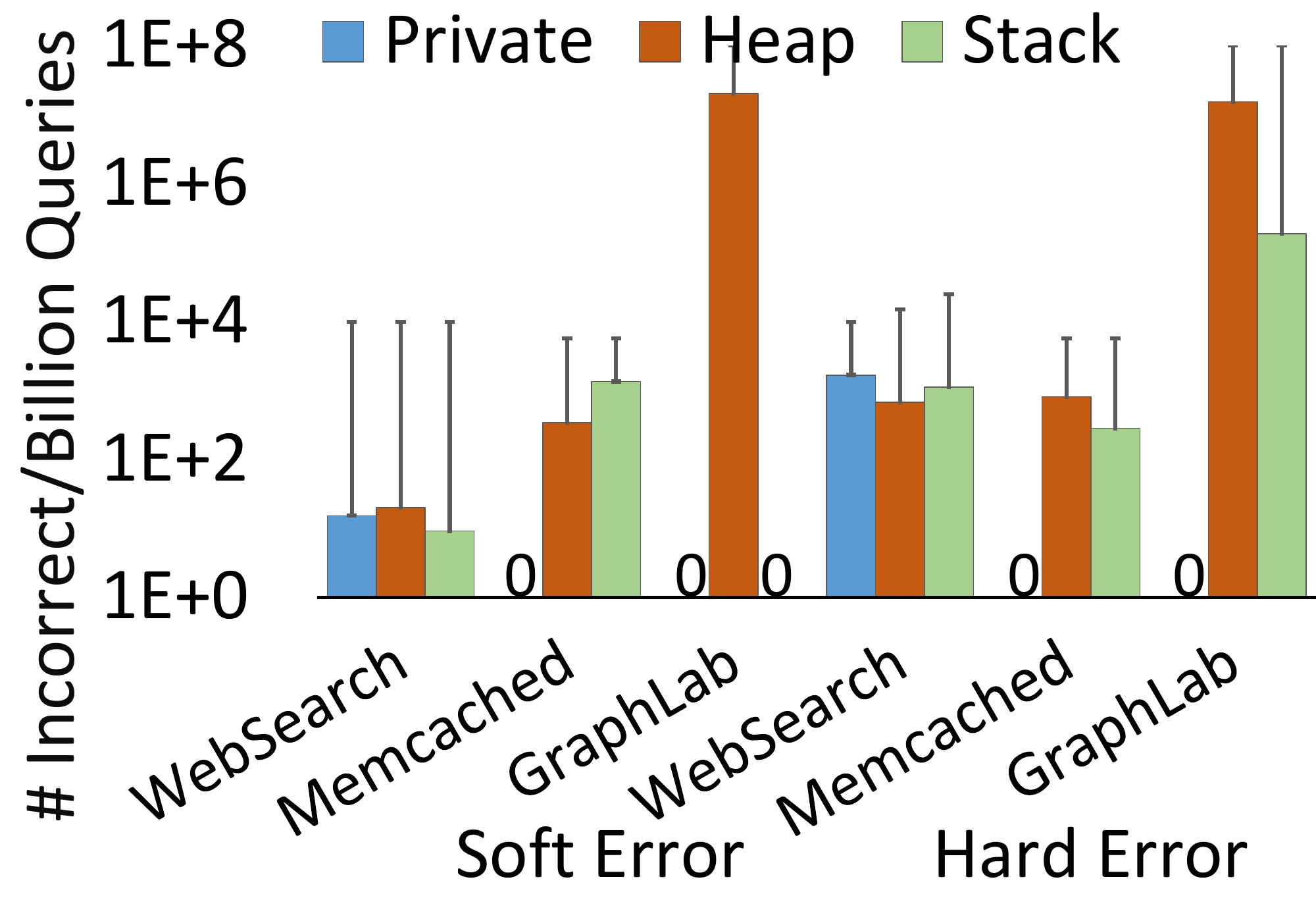}
\caption{Memory region incorrectness}
\end{subfigure}
  \caption{Memory region variations in vulnerability to single-bit soft and hard memory errors for the applications in terms of (a) probability of crash and (b) frequency of incorrect results. Reproduced from~\cite{luodsn14}.}
  \label{fig:intra-app-vulnerability}
\end{figure}

{\bf {Finding 2: Error Tolerance Varies Within an Application.}}
Figure~\ref{fig:intra-app-vulnerability}(a) plots the probability of each of the
three applications crashing due to the occurrence of single-bit soft or hard
errors in \emph{different regions} of their memory \chI{address space}.
Figure~\ref{fig:intra-app-vulnerability}(b) plots the rate of incorrect results
per billion queries under the same conditions, \chI{for cases where a crash did not occur}.

We make two observations from Figure~\ref{fig:intra-app-vulnerability}.
First, for some memory regions, the probability of an error leading to a crash
is much lower than for others (for example, in WebSearch, the probability of a
hard error leading to a crash in the \textit{heap} or \textit{private} memory
regions is much lower than in the \textit{stack} memory region).  Second, even
in the presence of memory errors, some regions of some applications are still
able to tolerate memory errors (perhaps at reduced correctness).  This may be
acceptable for applications such as WebSearch that aggregate results from
several servers before presenting them to the user, in which case the likelihood
of the user being exposed to an error is much lower than the reported
probabilities.  We therefore conclude that {\em memory reliability techniques
that treat all memory regions within an application similarly are inefficient
because there exists significant variance in the error tolerance among different
memory regions.}

\subsection{Other Findings}
\label{sec:char:other}

In Section~V-B of our DSN 2014 paper~\cite{luodsn14},
we discuss four other findings that we make based on our characterization data.
These findings focus on the memory error tolerance of WebSearch,
which we find to be representative of the behavior of all three of our
characterized applications.  In particular, we find that: 

\begin{itemize}[leftmargin=1.2em]

  \item 
    \chI{More severe} failures (i.e., failures that lead to system downtimes) due to
    memory errors tend to crash the application or
    system quickly, while \chI{less severe} failures tend to generate incorrect
    results periodically.

  \item Some memory regions are safer than others. 
    This indicates \chI{that either an application's access
    pattern or computational operations on different memory regions} can be the
    dominant factor to mask a majority of memory errors.

  \item \chI{More severe} errors mainly \chI{\emph{decrease}} correctness, as opposed to increase
    an application's probability of crashing.

  \item Data recoverability varies across memory regions.
    For data-intensive applications like WebSearch, software-only memory error
    tolerance techniques are a promising direction for enabling reliable system
    designs.

\end{itemize}


\section{Heterogeneous-Reliability Memory}

Based on the findings from our experimental characterization,
we propose {\bf \em heterogeneous-reliability memory} (HRM), a \chI{software/hardware cooperative framework} that
employs different levels of memory reliability within a single main memory subsystem to
optimize datacenter cost based on the memory error tolerance \chI{level} of applications \chI{and their memory regions}. We examine three dimensions, and their
benefits and trade-offs in the design space, for systems with heterogeneous
reliability \chI{memory}:  (1) hardware techniques to detect and correct errors, (2) software
responses to errors, and (3) the granularity at which different techniques are
used. Table~\ref{tbl:design-space} lists the techniques we considered
in each of the dimensions along with their potential benefits and trade-offs.

Using WebSearch as an example application, we evaluate and compare five \chI{example} design
points (three non-HRM systems, and two HRM systems):
\begin{itemize}[leftmargin=1.2em]

  \item {\bf \textit{Typical Server} (non-HRM):}  A baseline configuration resembling a typical
    server deployed in a modern datacenter.  All memory is homogeneously
    protected using SEC-DED ECC.

  \item {\bf \textit{Consumer PC} (non-HRM):}  Consumer PCs typically have no hardware
    protection against memory errors, reducing both their cost and reliability.

  \item {\bf \textit{Detect\&Recover} (\chI{HRM}):}  Based on our observation that some
    memory regions are safer than others, we consider
    an HRM system design that, for the \textit{private} region, uses parity in
    hardware to detect errors and responds by correcting them with a clean copy
    of data from disk in software (Par+R, parity and recovery), and uses neither
    error detection nor correction for the rest of its data.

  \item {\bf \textit{Less-Tested} (L; non-HRM):}  Testing increases both the cost and
    average reliability of memory devices~\cite{mutlu.date17,mutlu.imw13,patel.isca17}.  This system examines the
    implications of using less-thoroughly-tested memory throughout the \chI{\emph{entire} memory}
    system.

  \item {\bf \textit{Detect\&Recover/L} (HRM):}  This system evaluates the
    Detect\&Recover design with less-tested memory.  ECC is used in the
    \textit{private} region and Par+R in the \textit{heap} to compensate for the
    reduced reliability of the less-tested memory.

\end{itemize}
Section~VI-A of our DSN 2014 paper~\cite{luodsn14} discusses (1) the
metrics we use to evaluate the benefits and costs of the designs, and (2) the
memory error model we use to examine the effectiveness of the five designs.
\chI{We refer the reader to Section~VI-A in~\cite{luodsn14} for detail and a
full understanding.}

\begin{figure}[h]
\centering
\includegraphics[trim=00mm 255mm 0mm 0mm,clip,width=0.48\textwidth]{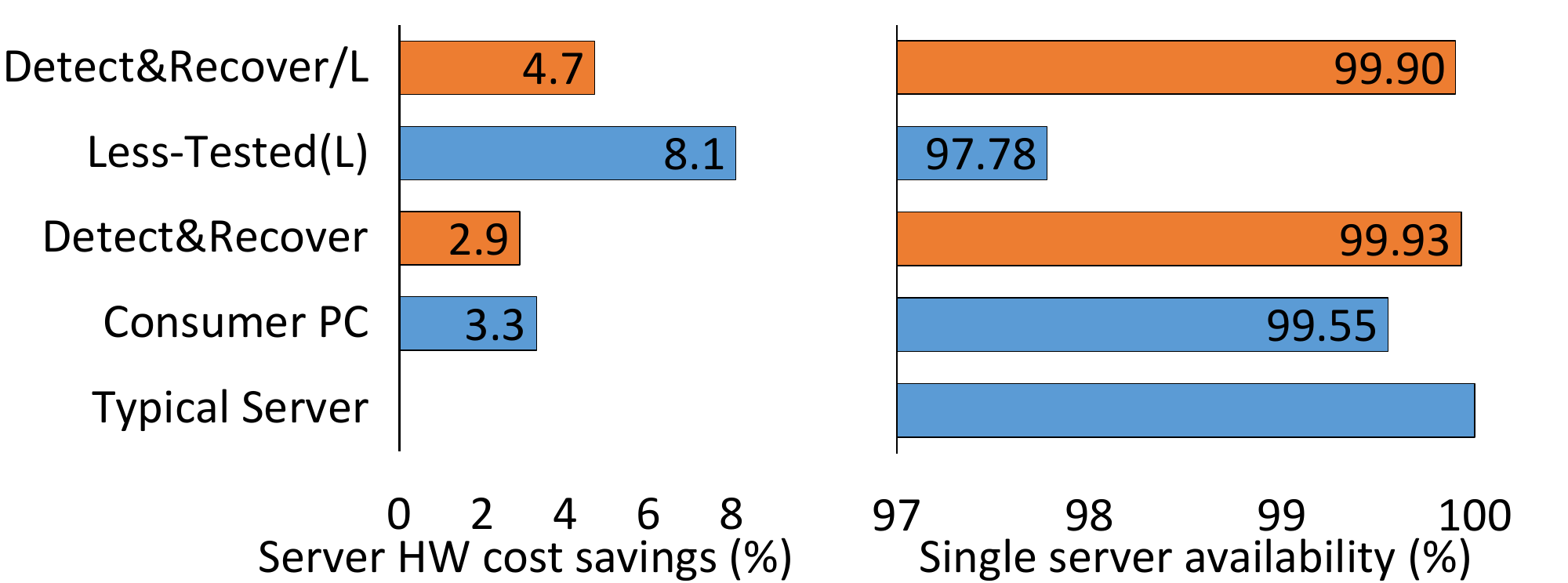}
\caption{
  Comparison of server hardware cost savings and single server availability for
  the five design points. Results extracted from~\cite{luodsn14}.
  \chI{Orange bars indicate HRM designs.}
}
\label{fig:avail}
\end{figure}

\chI{Our evaluation illustrates the inefficiencies of traditional homogeneous approaches
to memory system reliability, as well as the benefits of
heterogeneous-reliability memory system designs.}
%
\chI{Figure~\ref{fig:avail} shows the cost savings and 
\chII{single server availability} for our five evaluated design points.  We observe
from the figure} that the two highlighted \chI{example HRM}
design points (in orange color), which leverage our {\em heterogeneous-reliability} memory system
design, \chII{both can} achieve \chI{our} target single server availability of 99.90\% while
reducing server hardware cost by 2.9\% and 4.7\% respectively. We therefore
conclude that {\em heterogeneous-reliability memory system designs can enable
systems to achieve both high cost savings and high single server
availability/reliability at the same time}.

Section~VI of our DSN 2014 paper~\cite{luodsn14} contains a detailed analysis
of HRM, including
(1)~memory cost savings 
(Section~VI-B of~\cite{luodsn14}),
(2)~the expected crash and incorrect query frequency for each configuration
(Section~VI-B of~\cite{luodsn14}),
(3)~the maximum number of tolerable errors per month for each
application to achieve a reliability target (Section~VI-B of~\cite{luodsn14}), and 
(4)~a discussion of hardware/software support for and feasibility of HRM 
(Section~VI-C of~\cite{luodsn14}).
\chI{We summarize the key empirical findings here:
\begin{itemize}[leftmargin=1.2em]

\item Our two example HRM designs, Detect\&Recover and Detect\&Recover/L,
reduce memory costs by 9.7\% and 15.5\%, respectively, \chII{compared to the 
cost of the Typical Server system, which does not use HRM}.

\item The two example HRM designs limit the number of crashes to 3 and 4 per
server per month, respectively, and limit the incorrect query frequency to 9
and 12 per million queries, respectively.

\item Without \emph{any} error detection/correction, two out of our three evaluated applications
(WebSearch and Memcached) are able to achieve 99.00\% single server availability.

\end{itemize}

We therefore conclude that heterogeneous-reliability memory
system designs can enable systems to achieve both high cost
savings and high single server availability/reliability at the
same time. We believe that there is significant
opportunity in \chII{many data-intensive} applications for reducing
server hardware cost while achieving high single server
availability/reliability using our heterogeneous-reliability design
methodology.}


\section{Related Work} 

To our knowledge, our DSN 2014 paper~\cite{luodsn14} is the first to
(1)~perform a
comprehensive analysis of memory error vulnerability {\em for data-intensive
datacenter applications across a range of different memory error types;}
\chI{\chII{(2)~}propose the idea of \emph{heterogeneous reliability memory}, which
consists of multiple memory types with different levels of reliability and error
handling mechanisms; and 
(3)}~evaluate the cost-effectiveness of different {\em heterogeneous-reliability
memory organizations} with hardware/software cooperation.
We discuss related research in memory error vulnerability and DRAM
architecture below, categorizing the works into six broad classes: (1)~memory errors in datacenters, 
(2)~characterizing application error tolerance, (3)~hardware-based memory
reliability techniques, (4)~software-based memory reliability techniques, 
(5)~exploiting application error tolerance, and (6)~heterogeneous (hybrid)
memory architectures.


\chI{\textbf{Studies of Memory Errors.}
Various works~\cite{Schroeder09, meza.dsn15, Li10,
  Sridharan2012, Sridharan2013, sridharan.asplos15}
   have conducted studies of DRAM error
rates that are deployed in production datacenters, 
studying failures across a large sample
size.
These works note that memory errors occur frequently
in datacenters, and are induced by a number of error
sources.
In particular, one of these studies empirically demonstrates the increased
memory errors \chI{and increased} memory cost to tolerate these
errors in large-scale datacenters~\cite{meza.dsn15}.
A recent work~\cite{gottscho.cal17} examines how various hardware
and software techniques to detect and mitigate errors introduce
significant performance degradation in production datacenters.
This work shows that for WebSearch, software error handling techniques can
induce a performance overhead of 3746$\times$~\cite{gottscho.cal17}.
These studies motivate the need for a low-overhead, cost-effective approach
to memory reliability, and motivate us to further explore hardware--software
cooperative techniques such as HRM.}

\chI{There are several studies that characterize
various sources of errors in DRAM at a fine granularity. 
Many of these works observe how specific factors affect
DRAM errors, analyzing the impact of
temperature~\cite{el-sayed-sigmetrics2012,lee.hpca15} and hard
errors~\cite{hwang.asplos12}. A large number of works study errors through controlled experiments,
usually using FPGA-based DRAM testing infrastructures like SoftMC~\cite{hassan.hpca17},
to investigate errors due to
retention time~\cite{liu2012raidr, khan.cal16,khan.micro17,liu2013experimental,
    khan.sigmetrics14, khan.dsn16, qureshi2015avatar, patel.isca17,
    hassan.hpca17}, disturbance from neighboring DRAM
cells~\cite{mutlu.date17,kim.isca14,kim.thesis15,jung-memsys2016}, latency variation across/within DRAM
chips~\cite{lee.hpca15, lee.thesis16,chandrasekar.date14, chang.sigmetrics16,
  lee.sigmetrics17,chang.thesis17}, and supply voltage~\cite{chang.sigmetrics17,chang.thesis17}. None of
these works study memory errors in a system with heterogeneous-reliability memory.}

\textbf{Classifying Application Error Tolerance.} \chI{Error} injection
techniques based on hardware watchpoints~\cite{Li10, Messer04}, binary
instrumentation~\cite{Li12}, and architectural simulation~\cite{xuanhua} have
been used to investigate the \chI{impact} of memory errors on application behavior,
including execution times, application/system crashes, and output correctness.
These works study a range of applications including SPEC CPU benchmarks,
web servers, databases, and scientific applications. In general, these works
conclude that not all memory errors cause application/system crashes and \chI{many memory errors} can be
tolerated with minimal difference in \chI{the application} outputs.  We generalize this
observation to data-intensive applications, and leverage it to reduce datacenter
TCO\@.  Recent work~\cite{subasi.ccgrid17} develops a \chI{Markov-chain} model for the error tolerance of
HPC applications.
Approximate computing techniques~\cite{approx12, Bornholt14, iACT}, where
the precision of program output can be relaxed to achieve better performance or
energy efficiency, offer further opportunities for leveraging the
error-tolerance of application data, though these typically require \chI{very careful} changes to
the program source code.

\textbf{Hardware-Based Memory Reliability Techniques.}  There are
various ECC techniques for memory, and we list the most dominant ones in
Table~\ref{tbl:mem-rel-cost}.  Using eight bits, SEC-DED can correct a single
bit flip and detect up to two bit flips out of every 64~bits.  DEC-TED is a
generalization of SEC-DED that uses fourteen bits to correct two and detect
three flipped bits out of every 64~bits.  Chipkill~\cite{Dell97} improves reliability by
interleaving error detection and correction data among multiple DRAM
chips.  RAIM~\cite{Meaney12} is able to tolerate entire DIMMs
failing by storing detection and correction data across multiple DIMMs.
Virtualized ECC~\cite{Yoon2010} maps ECC to software-visible locations in
memory so that software can decide what ECC protection to use.  While
Virtualized ECC can help reduce the DRAM hardware cost of memory reliability,
it requires modification to the processor's memory management unit and cache(s).

Recent works propose new hardware-based techniques to tolerate soft and hard
memory errors efficiently. \chI{We \chII{break these down} into four categories:}
(1)~\chI{\emph{Tolerating soft errors:}} BambooECC~\cite{kim.hpca15} proposes a
new single-tier ECC family that enables adaptive graceful downgrade of ECC
capabilities. CleanECC~\cite{gong.micro15} provides both high memory reliability
and flexible memory access granularity by using fine-grained error detection
and coarse-grained error correction. XED~\cite{nair.isca16} uses in-DRAM ECC
to reduce the overhead of double Chipkill. (2)~\chI{\emph{Tolerating hard
errors:}} ArchShield~\cite{nair.micro13} proposes an architectural framework to identify and
tolerate hard errors caused by DRAM cell failures. Citadel~\cite{nair.taco16}
proposes to tolerate large-granularity failures, such as row/bank failures, by
replacing them with spares. Other works propose to
identify and mitigate potentially recurring memory errors by page
offlining~\cite{meza.dsn15}, online testing~\cite{khan.sigmetrics14, patel.isca17},
and multi-rate refresh\chI{~\cite{qureshi2015avatar,liu2012raidr}}.
(3)~\chI{\emph{Reducing memory cost:}}
FrugalECC~\cite{kim.sc15} proposes a new flexible granularity compression to reduce the
redundancy and energy consumption of ECC. Morphable ECC~\cite{chou.dsn15} proposes to
reduce DRAM refresh overhead by reducing ECC strength to 6-bit ECC when the
DRAM is in idle mode. (4)~\chI{\emph{End-to-end memory error protection:}}
AIECC~\cite{kim.isca16} provides end-to-end protection for clock, control, command, and address 
(CCCA) signals in addition to data signals.

\textbf{Software-Based Memory Reliability Techniques.} Previous works
(e.g.,\chII{~\cite{hwang.asplos12, Tang2006, Schirmeier2012, meza.dsn15}}) show that the OS
retiring memory pages after a certain number of errors can eliminate up to 96.8\%
of detected memory errors.  While these techniques improve system
reliability, they still require costly ECC hardware for detecting and identifying
memory pages with errors.  Other works attempt to reduce the impact of memory
errors on system reliability by writing more reliable
software~\cite{Borchert2013}, modifying the OS memory allocator~\cite{Samurai},
or using a compiler to generate a more error-tolerant version of the
program~\cite{Chang06, Benso00}. Other algorithmic solutions (e.g., memory
bounds checks~\cite{hyungmin}, watchdog timers~\cite{hyungmin}, and checkpoint
recovery\chI{~\cite{LiDSN08, Xu12, Li08,li.date08,li.vts10,li.iccad09,wang.ftcs95,
constantinides.micro07,constantinides.micro08,constantinides.tc09}}) can also be used to \chI{improve}
resilience to memory errors.

Li et al.~\cite{li.memsys16} propose to deploy software-based ECC
in an in-memory key-value store, and show that it incurs low performance
overhead. Recent works~\cite{subasi.ccgrid17, zheng.isca17} \chI{improve} upon traditional RAIM-3 and
use selective replication to reduce unnecessary memory
redundancy. SDECC~\cite{gottscho2016software,
gottscho2017low} proposes to use strong
error detection in the hardware, while tolerating hard memory errors and
recovering from soft errors in the software.

\textbf{Exploiting Application Error Tolerance.} Flikker~\cite{song} proposes a
technique to trade off DRAM reliability for energy savings. It relies
on the programmer to separate application data into vulnerable or tolerant data.
Less reliable mobile DIMMs have been proposed~\cite{boom12, horowitz12}
as a replacement for ECC DIMMs in servers to improve energy efficiency.
Recent work~\cite{ongaro.sosp11} shows that RAMCloud can recover 35~GB of data from a failed server
in 1.6~seconds using a log-structure storage.

\textbf{Heterogeneous (Hybrid) Memory Architectures.} \chI{Various recent} works
(e.g.,\chII{~\cite{Moin2009, Yoon2012, phadke2011, rajeev2012, meza2012, li.cluster17,
meza2013case, yu.micro17, ramos.ics11, zhang.pact09, agarwal.asplos17, 
dulloor.eurosys16, pena.cluster14, bock.iccd16, gai.hpcc16, liu.iccd16}}) explore
the use of heterogeneous memory architectures, consisting of multiple different
types of memories. These works are mainly concerned with either mitigating the
overheads of emerging memory technologies or improving performance and power
efficiency. They do not investigate the use of multiple devices with different
\chI{\emph{error correction}} capabilities.
CREAM~\cite{luo2017using} and Odd-ECC~\cite{malek2017odd} develop low-cost 
techniques to provide flexible provisioning
of memory error correction capabilities.
Recent works~\cite{arslan2015performance,
wang2015heterogeneous} apply our heterogeneous reliability idea to processor caches to
achieve better cost-reliability trade-offs.




\section{Significance and Long-Term Impact}

We believe that our DSN 2014 paper~\cite{luodsn14} will have long-term impact for three \chI{major}
reasons. First, it emphasizes and aims to solve the increasing cost of ensuring memory
reliability as the error rates of memory devices continue to grow, \chI{which is a major trend as memory
technology scales to smaller technology nodes~\cite{mutlu.date17,mutlu.imw13}}.
Second,
it tackles memory system cost in datacenters,
which is a problem that we expect will be
increasingly important in the future.
Third, it proposes a \chI{novel framework} that
uses hardware--software co-design to improve memory system reliability \chI{as well as cost}, thereby
hopefully inspiring future works to exploit software characteristics to improve
system reliability and reduce system cost \chI{(and other important metrics)}.

{\bf Increasing Memory Error Rate.} As DRAM scales to smaller process technology
nodes, the reliability of DRAM continues to degrade\chI{~\cite{kang2014co,
mutlu2013, meza.dsn15, Sridharan2013, Sridharan2012, Schroeder09, sridharan.asplos15,
hwang.asplos12, mutlu.superfi14,mutlu.imw13, mutlu.date17}}.  For example, recent works 1)~show the existence of
disturbance errors in commodity DRAM chips operating in the
field\chI{~\cite{kim.isca14, mutlu.date17}}; 2)~experimentally demonstrate the increasing
importance of retention-related failures in modern DRAM
devices\chI{~\cite{liu2012raidr, liu2013experimental, khan.sigmetrics14,
hassan.hpca17, khan.dsn16,
patel.isca17, khan.cal16, khan.micro17,
qureshi2015avatar, patel.isca17}}; 3)~examine the trade-off between DRAM reliability and
latency~\cite{lee.hpca15, chang.sigmetrics16, lee.sigmetrics17,
chandrasekar.date14, hassan.hpca17, chang.thesis17, lee.thesis16,
chang.sigmetrics17, kim.hpca18};
and 4)~advocate, \chI{including} in a paper co-written by the Samsung and Intel memory
design teams~\cite{kang2014co}, for the use of in-DRAM error correcting codes to overcome the
reliability challenges~\cite{kang2014co, qureshi2015avatar}. As a result of
decreasing DRAM reliability, maintaining the effective error rate at the levels we have
today can (1)~increase DRAM cost due to decreased yield, expensive quality
assurance tests, and/or extra capacity for storing stronger error-correcting codes;
or (2)~reduce DRAM performance due to frequent error correction and logging. All of these solutions
\chI{might} make DRAM \chI{technology} scaling more difficult and less appealing\chI{~\cite{mutlu.imw13,mutlu2013,mutlu.date17}}.  Our paper proposes
a solution that enables the use of DRAM with \emph{higher} error rates while still achieving
reasonable application reliability, which can enable \chI{much} more efficient scaling of
DRAM to smaller technology nodes in the future.

Other memory technologies such as NAND flash memory~\cite{cai2014neighbor, 
cai2013threshold, cai2013program, cai2012error, cai2012flash,
cai2013error, meza2015sigmetrics, cai.hpca15, cai.dsn15, cai.hpca17,
luo.jsac16, luo.hpca18, cai.procieee17, cai.procieee.arxiv17, cai.arxiv17, fukami.di17}, phase-change memory
(PCM)\chI{~\cite{lee2009architecting, lee2010phase, lee2010technology, Moin2009,
meza2013case,meza.tr12}}
and STT-MRAM~\cite{kultursay2013evaluating, meza2013case} also
show a similar decreasing trend in their reliability with process technology
scaling and the advent of multi-level cell (MLC)
technology~\cite{mutlu.imw13}.  For example, like DRAM, NAND flash memory
suffers from retention errors~\cite{cai2012error, cai2012flash, cai.hpca15, cai.procieee17, cai.procieee.arxiv17, cai.arxiv17}\chI{,}
cell-to-cell program interference errors~\cite{cai2012error,
cai2013program, cai2014neighbor, cai.procieee17, cai.procieee.arxiv17, cai.arxiv17}, \chI{and read disturb
errors~\cite{cai2012error, cai.dsn15, cai.procieee17, cai.procieee.arxiv17, cai.arxiv17}}.  Additionally, NAND flash memory suffers from
program/erase cycling errors~\cite{cai2012error, cai2013threshold}, \chI{and} programming
errors~\cite{cai.hpca17, parnell.globecom14, luo.jsac16}. PCM suffers from endurance
issues~\cite{lee2009architecting, Moin2009, 
lee2010technology} and resistance drift~\cite{ielmini2007physical}. 
HRM can be applied to these memory technologies
with slight modifications to enable reliable high-density non-volatile devices
in the future.

{\bf Increasing Datacenter Cost.}
Recent studies have shown that capital costs can account for the majority (e.g.,
around 57\% in~\cite{Barroso09}) of datacenter TCO (total cost of ownership). As
part of the cost of a server, the cost of the memory is comparable to that of
the processors\chI{~\cite{Kozyrakis10}}, and is likely to exceed processor cost and become the dominant
cost for servers running data-intensive applications such as web search and
social media services\chI{~\cite{Reddi10,memcached,graphlab}}. As future datacenters grow in scale, datacenter TCO will
become an increasingly important factor in system design.  Our paper
demonstrates a way of optimizing datacenter TCO by reducing the cost of the
memory system. The cost savings can be significant due to the increasing scale
of such datacenters~\cite{zdnet}, making our proposed technique hopefully more
important in the future.

{\bf Hardware--Software Co-Design.} Our solution, heterogeneous-reliability
memory, utilizes hardware--software cooperative design to reduce system cost. Our DSN 2014 paper~\cite{luodsn14}
demonstrates the benefits of exploiting application characteristics to
improve overall system design. For example, it shows that a significant number
of errors can be corrected in software by reloading a clean copy of the data
from storage. This motivates us to rethink the placement of different
functionalities (such as error detection and error correction) \chI{across
different \chII{system} components and across software versus hardware} to improve the
cost--reliability trade-off.


Our DSN 2014 paper~\cite{luodsn14} has started a community
discussion~\cite{zdnet} on the feasibility of solving the problem of memory
reliability by exploiting application memory error tolerance in the future,
inspiring reporters to ask the question: ``How good does memory need to be?''
We hope that our characterization results and mechanisms will hopefully continue to inspire future works that can provide
efficient and extensive characterization/estimation of application-level memory
error tolerance~\cite{Foutris2014}, which can make our proposed technique
applicable to a broader set of applications.

\chI{Two example works that build on ours include Odd-ECC~\cite{malek2017odd}
and CREAM~\cite{luo2017using}. Odd-ECC provides a mechanism to enable different
levels of fault tolerance for the data stored in a commodity DRAM \chII{module}. Odd-ECC maps
the ECC bits to a memory address aligned with the data so that the memory controller
can access both the data and the ECC bits efficiently. CREAM provides a mechanism
to dynamically adjust the tradeoff between memory capacity/bandwidth used for ECC bits
and fault tolerance within an ECC DRAM module. CREAM proposes several data
layouts that reduce page faults and improve memory performance significantly when strong fault
tolerance is not needed.}




%

\section{Conclusion}
\label{sec:conc}

In our DSN 2014 paper~\cite{luodsn14}, we \chI{develop} a new methodology to quantify the tolerance of
applications to memory errors.  Using this methodology, we \chI{perform} a case
study of three new data-intensive workloads that \chI{show}, among other new
insights, that there exists a diverse spectrum of memory error tolerance both
within and \chI{across} these applications. \chI{Based on this observation,
we introduce the idea of heterogeneous-reliability memory (HRM), which combines
multiple different memories that have different reliability characteristics and
error correction capabilities.}  We \chI{propose} new hardware/software
heterogeneous-reliability memory system designs, and \chI{evaluate} them to show that
(1)~the one-size-fits-all approach to reliability in modern servers is inefficient
in terms of cost, and (2)~heterogeneous-reliability systems can achieve the
benefits of both low cost and high single server availability/reliability.  We
hope that our techniques can enable the use of lower-cost memory devices to
reduce the server hardware cost of datacenters, and that our analyses will spur
future research on heterogeneous-reliability memory systems. As DRAM technology
scales into small feature sizes and becomes less reliable and memory
cost becomes more important in datacenters in the future, we hope that our
findings and ideas will inspire more research to improve the cost--reliability
trade-off in memory systems.
\chI{We believe \chII{different HRM designs can be employed to optimize
other key trade-offs and target metrics (e.g., performance vs.\ energy 
consumption) in modern systems.}
Our DSN 2014 paper just scratches the surface of
a large amount of research and design space to be explored.}


\section*{Acknowledgments}

\chI{We thank Saugata Ghose for his dedicated effort in the preparation
of this article.}
We thank the anonymous reviewers and the members of SAFARI research group for
feedback. We acknowledge the support of Microsoft and Samsung. This research was
partially supported by the Intel Science and Technology Center for Cloud
Computing and the NSF (grants 0953246, 1065112, and 1212962).


\bibliographystyle{IEEEtranS}
\bibliography{ref}

\end{document}